\documentclass[11pt]{article}

\usepackage[preprint]{acl}

\usepackage{times}
\usepackage{comment}
\usepackage{latexsym}
\usepackage{amsmath}
\usepackage{bbm}
\usepackage{amsfonts}

\usepackage[T1]{fontenc}

\usepackage[utf8]{inputenc}

\usepackage{microtype}

\usepackage{inconsolata}

\usepackage{graphicx}

\usepackage{booktabs}
\usepackage{xcolor}
\usepackage{colortbl}
\usepackage{array}
\usepackage{pifont}  
 
\definecolor{checkbg}{RGB}{225, 240, 225}    
\definecolor{partialbg}{RGB}{250, 240, 215}  
\definecolor{xbg}{RGB}{248, 225, 225}        
 
\newcommand{\yes}{\cellcolor{checkbg}\ding{52}}
\newcommand{\prt}{Maybe}
\newcommand{\no}{\cellcolor{xbg}\ding{55}}

\newcommand{\igl}[1]{\textcolor{blue}{\textbf{[igl: #1]}}}

\title{Rigorous Interpretation Is a Form of Evaluation}

\author{%
Isabelle Lee\thanks{Correspond to \texttt{lee.isabelle.g@gmail.com}}\\
USC\\
\And
Emmy Liu\\
CMU\\
\And 
Cathy Jiao\\
CMU\\
\And 
Brihi Joshi\\
USC\\
\AND
Dani Yogatama\\
USC\\
\And
Fazl Barez\\
Oxford \& WhiteBox\\
\And 
Michael Saxon\\
UW\thanks{Now at Google Deepmind.}\\
}

\begin{document}
\maketitle
\begin{abstract}

Current machine learning models are evaluated through behavioral snapshots, with benchmark accuracies, win rates and outcome-based metrics. 
Model explanations and evaluations, however, are fundamentally intertwined: understanding why a model produces a behavior can be as important as measuring what it produces. 
If we trusted interpretability, we argue that it can serve not merely as diagnostics but as a richer and more principled form of model evaluation beyond surface-level performance metrics.
We explore three ways interpretability can function evaluatively: (1) fixing problems by identifying the root causes of unwanted behavior, (2) detecting subtly faulty mechanisms that invalidate model outputs, and (3) predicting potential issues before they arise by fully understanding the model’s weaknesses.
To fulfill its evaluative potential, we argue that interpretability methods must generate claims that are falsifiable, reproducible, and predictive---that is, interpretability must meet scientific standards.

\end{abstract}

\section{Introduction}

Current machine learning models are evaluated on their outputs. 
Typically, they are behaviorally assessed by performances on held-out data, win rates on benchmarks---some aggregate of these task metric scores. 
While these forms of evaluations capture immediate summary of behavior in a few numbers, they only capture surface competence. 
Two models may achieve identical behavior while relying on radically different internal mechanisms for example: one grounded in sound reasoning and computation and therefore structurally robust, the other brittle, heuristic driven.\looseness-1

This position paper argues that \textit{interpretability} has the potential to expand current model assessment beyond surface-level behavioral evaluation to mechanism-level scrutiny. 
This is not to suggest that benchmarks are uninformative; rather, interpretability can deepen and extend them. 
However, for interpretability to serve this evaluative role, it must meet scientific standards. 
Interpretability claims must be \textit{falsifiable, reproducible, and predictive.} Without these properties, interpretability remains descriptive rather than evaluative.\looseness-1

If interpretability satisfies these standards, it can transform evaluation from a retrospective summary of behavior into a dynamic, mechanism-sensitive process.
By making internal structures accessible, interpretability enables: (1) causal diagnosis and repair of known failures, (2) detection of faulty reasoning even when outputs appear correct, and (3) anticipation of failures before they manifest behaviorally. More specifically:\looseness-1

\begin{enumerate}

    \item \textbf{Falsifiability enables debugging.} 
    In this scenario, we identified undesirable model behavior using well-defined evaluation criterion, and we want to use interpretability to debug and fix it. Falsifiability, defined as the quality of a claim being structured so it can in principle be proven wrong through targeted tests or interventions, is essential here. If we can make falsifiable, causal interpretability claims, we can reliably use model-based approach to correct unwanted associations at an architectural level, thereby addressing the root cause of the problem. Static benchmarks may reveal when models fail and sometimes characterize the nature of those failures, but they do not explain why the failure occurs or how to fix it. Interpretability may offer a pathway for more principled debugging.

    \item \textbf{Reproducibility ensures detection of faulty mechanisms.} 
    In this scenario, the evaluative criterion is unknown or poorly defined, so we use interpretability to identify the reasoning or associations behind a model decision. 
    Interpretability here functions as a \textit{mechanism-level evaluation}, because it can assess internal reasoning rather than just output, which can detect when a model produces a plausible output while relying on spurious correlation or heuristic shortcuts---that is, \textit{if our interpretability methods met the standards of robustness, reliability, and reproducibility.}
    To serve this role as a scientific tool, interpretability methods must be reproducible and reliable, consistently identifying the same mechanisms across runs and remaining robust to perturbations; otherwise, apparent findings may reflect noise rather than genuine faulty reasoning.

    \item \textbf{Predicting future failures.} 
    In this scenario, we don’t have a specific pre-set evaluative criterion or particular decision we wish to examine. Instead, we aim to predict potential problematic behavior or criteria that the model might struggle with in general. By applying the framework of \textit{scientific prediction} as testing of our understanding, we generate interpretability claims that are predictive. By analyzing the model’s internal behavior on a natural or in-distribution input set, interpretability techniques can help generate challenge sets or anticipate which types of inputs might cause issues. Practically, this can be seen as a way to stress-test the model before it is released to the public, allowing further refinement.
    
\end{enumerate}

Taken together, interpretability can be reframed not merely as an exploratory add-on, but as a scientifically grounded and reliable explanation of behavior: an evaluation in its own right. 
Current evaluations are largely a retrospective summary of behaviors, but interpretability has the potential to transform evaluations into a forward-looking, mechanism sensitive discipline. 
When held to rigorous standards, interpretability can be evaluation.

\section{Falsifiability enables debugging}
\label{sec:fixing-unwanted-behavior}

One direct application of interpretability is to debug a model by identifying the internal origins of a model's problematic output and correcting them by direct modification or retraining. 
Unlike evaluation, interpretability can not only characterize the problematic behavior, but also may expose the internal failures underlying a particular behavior.
However, for the interpretability claims to be testable and therefore reliably correctable, they must be falsifiable \cite{popper2005logic, leavitt2020falsifiableinterpretabilityresearch}---that is, they can in principle be counterfactually \textit{disprovable}.

\citeauthor{leavitt2020falsifiableinterpretabilityresearch} critique a tendency in parts of the interpretability literature to prioritize intuition-building over the formulation of falsifiable hypotheses. 
In particular, they argue that attention heatmaps, component visualizations, and similar input-based explanations often function as descriptive aids rather than as testable claims about a model’s internal mechanisms. Such approaches may help researchers build intuitions about how models behave, but they do not necessarily generate hypotheses that can be rigorously confirmed or disconfirmed. More broadly, they caution against explanatory claims that merely redescribe behavior without exposing themselves to potential refutation. In this respect, causal and mechanistic interpretability approaches appear to better meet \citeauthor{leavitt2020falsifiableinterpretabilityresearch}'s standard of hypotheses strength, as they more frequently articulate hypotheses that range from weakly to moderately falsifiable.

Current causal interpretations, however, face structural limitations that undermine this falsifiable debugging ideal. Many proposed features are \textbf{non-specific}: intervening on them affects multiple behaviors rather than isolating a single target function, leaving their representational role unclear. They are often \textbf{inconsistent}: the same intervention does not reliably produce the same behavioral shift across similar inputs. And they are subject to \textbf{underdetermination}: multiple distinct features can appear to influence the same behavior, with no principled way to identify which, if any, constitutes the true causal mechanism. When these conditions obtain, the debugging analogy collapses---the “feature” no longer behaves like a stable component but like a correlated direction in a high-dimensional space. A genuinely falsifiable debugging claim would require a stable causal unit, predictable intervention effects, and clear criteria for success or failure. Many current mechanistic interpretations fall short of these standards.

In this section, we first consider cases where interpretability methods do succeed in identifying meaningful, actionable features (\S\ref{sec:interp-sometimes-can-fix-problems}). We then turn to cases where interpretability, particularly causal methods, breaks down (\S\ref{sec:disconnect_interp_eval}), and examine the specific modes of failure in those settings.

\subsection{\textit{Sometimes} interpretation can fix problems}
\label{sec:interp-sometimes-can-fix-problems}
Interpretability, at its best, helps identify the root cause of model misbehavior by illuminating how predictions are made. 
For example, explanations produced by \textit{concept bottleneck} methods learn concepts specified by domain experts; these concepts are then used to constrain the model outputs, preventing errors at test time \citep{pmlr-v119-koh20a, Losch2019InterpretabilityBC}. Concept bottleneck interpreters are developed through \textit{supervised} training on explicit concept labels provided by domain experts.
These highlighted concepts outputted demonstrate what a model uses to arrive at a certain output.
To fix the incorrect model output, we can directly intervene on weights of the model corresponding to found explanations.
Concept bottleneck methods have shown promise in medical imaging; \citet{pmlr-v119-koh20a} reported significantly improved performance in modeling knee osteoarthritis from X-rays after domain experts edited the model's predictions.
However, this test-time debugging approach relies on experts' curated concepts, which are not easily transferable across tasks or, in some cases, even datasets \citep{Hayashi2023BayesianGE, Shin2023ACL}. \looseness-1

Another recent example is Sieve \citep{karvonen2024sieve}, a code-specific suite of SAEs trained on public coding datasets. By promoting LM activation features related to regular expressions, Sieve improved performance on examples that required regular expressions. As in this scenario, interpretability at its best can enable direct debugging through interventions that have an explainable effect on model behavior.

\subsection{Interpretation often can't fix problems}
\label{sec:disconnect_interp_eval}

In mechanistic interpretability, an interpretability method is seen as faithful if it highlights features \textit{causally} linked to an understood model behavior in output, as verified by interventions on model components or modifications to input data distributions \citep{Saphra2024Mechanistic, Mueller2024TheQF}. 
While these methods have shown promise, causal interpretations often fail to be falsifiable. They are underdetermined and inconsistent---often breaking down under distributional shifts, where the assumptions underpinning these causal relationships no longer hold. In this section, we examine one widely used mechanistic interpretation: causal interventions using features found with SAEs through steering. We then examine evidence of the brittleness of SAE feature interpretations. Such interventions cannot be reliably used in their current form for debugging. \looseness-1

\paragraph{Sparse Autoencoders} Sparse autoencoders (SAEs) are a popular method used to interpret learned concepts in models. An SAE functions as a regular autoencoder with additional constraints that encourage the intermediate dimension to have sparse activations \citep{ng2011sparse}. In practice, an SAE takes in an activation vector from a language model of size $N$, and sparsely reconstructs the original input activation in the hidden layer of size $M > N$. 
Recently, SAEs have been employed to analyze the activations of large language models by treating the hidden layer as a dictionary of concepts after training \citep{cunningham2023sparseautoencodershighlyinterpretable, gao2024scalingevaluatingsparseautoencoders}.\looseness-1

Because SAEs allow us to examine a dictionary of concepts learned by a model, we may potentially use this dictionary to examine how learned concepts interact. 
For example, we could see if the features for gender and profession tend to co-occur in certain ways, when debugging and correcting for gender bias. 
However, for these associations to be useful in debugging models, they must remain stable across contexts and distributions. 

\paragraph{Limitations of SAEs}
SAEs can suggest causal interventions through steering, or modifying a model's behavior by adjusting its activations in directions mapped to particular concepts. In practice, however, the effectiveness of steering can vary widely across different inputs, with the intervention failing or causing unintended consequences on specific examples \citep{tan2025analyzinggeneralizationreliabilitysteering}. It may be possible to improve on naive steering, for instance by trying to minimize the effects on other latent features 
when finding a steering vector as in \citep{chalnev2024improvingsteeringvectorstargeting}, but this does not fundamentally solve the problem of inconsistent feature interpretations in different contexts. 

Although there is some evidence that SAEs trained on different models may share some common concepts and patterns of feature organization \citep{lan_sparse_2024, Li2024TheGO}, at a more granular level the concepts recovered from SAEs depend on what data the SAE was trained on \citep{Paulo2025SparseAT}. 
An SAE is trained on activations from a layer of the target model being interpreted. But what inputs is the target model generating its activations from? Model behavior and activations may vary under different data distributions, and therefore SAE concepts found in one domain may not apply to others. 

In one instance \citep{SAEsAreHighlyDatasetDependent}, when training SAEs on the same model, a guardrail ``refusal'' feature \citep{arditi2024refusallanguagemodelsmediated} identified by activations from chat-formatted data failed to generalize when the target model was provided with webtext input instead. For example, suppose we trained an SAE on model activations using input data in which profession and gender were strongly correlated. If our analysis revealed a feature that linked gender and profession, we could apply an intervention to mitigate this bias. However, interventions based on this feature might fail because the same feature might have different functions, and the appropriate feature set may even differ, when the model is handling contexts where the presumed correlation between gender and profession no longer holds. 

\section{Reproducibility ensures detecting of faulty mechanisms}
\label{sec:detecting-problems}

We have now described how to analyze an output error in evaluation. What about cases where there is no observable error? Sometimes the model generates a seemingly plausible output, but relies on flawed internal mechanisms to produce that result. For example, we might need to address concerns of biased judgments on ambiguous inputs; determine whether a learned algorithm might fail on edge cases; or comply with regulations that ban incorporating protected characteristics for certain decisions. These concerns might not be apparent on individual outputs, prohibiting common metrics like accuracy, but they can affect behavior patterns across a distribution. We argue that \textit{reproducible interpretability} can aid with this ambiguity in evaluation, where perhaps the evaluative objective has not been clarified or is ambiguous by nature. 

\paragraph{An example: gender bias in machine translation} Let's consider an example where a model which translates Spanish to English: \textit{The doctor lost a stethoscope} as \textit{El doctor perdió un estetoscopio}. 
The English sentence does not provide any hints as to the gender of the doctor, so the model produces a valid translation which assumes the doctor is male. 
Because gender marking is mandatory in Spanish, the model has to assume some gender on the part of the doctor, and is perhaps relying on the linguistic default of masculine. 
However, the gender may have been assumed on the basis of the doctor's profession from the training data distribution, provoking concerns of model bias. Did the model choose \textit{el} based on the profession of \textit{doctor}? Perhaps we can answer this question using interpretability tools.

We hope to use interpretability to assess the following bias claim: ``this model chose a masculine token because it referred to a doctor.'' 
We would find that our interpretations are only useful for this claim if they are robust and reliable even when shifting away from the training data distribution.
For simplicity, let's assume a toy training data where all male referents are doctors and all doctors are male. 
Say we base the explanation on a specific neuron causally related to the masculine output token \textit{el}; this neuron's activation determines the output token to be masculine. To define the \textit{explanation distance}, then, a pair of inputs have similar explanations if they feature similar activation for this particular neuron. 

We seek to explain this neuron through a particular semantic relation between input (profession) and output (gender), such that the explanation distance is proportional to the semantic distance between two input and output pairs. 
To claim that the output gender depends on the referent being a doctor (and the model is therefore biased), we need this neuron to activate if and only if the referent is a doctor. 
Suppose that for every example similar to our toy training distribution, this condition holds---the activation is similar when the semantic relation between the profession and the referent is doctor-male.
However, we may also find that the neuron also activates when the relation instead simply identifies the referent's gender as male (i.e. ``\textit{the doctor is the man.}'' translates to ``\textit{El doctor es el hombre.}''). 
As a result, we can interpret the neuron as identifying a doctor referent or a male referent. 
The neuron's interpretation is underdetermined in the train set because the concepts ``doctor'' and ``male'' happen to be equivalent.  If we choose the former interpretation, the model appears to be deciding gender based on profession---a potentially undesirable bias. However, if we choose the latter interpretation, the neuron is directly expressing the gender of the referent---so its role in selecting output gender does not imply bias. 

In choosing a useful interpretation, we must aim for reproducibility: the interpretation should still hold for inputs out-of-distribution as well as for examples similar to the training distribution. If the robust, reliable, and therefore, reproducible interpretation defines the relation as referent profession rather than referent gender, we can use this neuron to detect biased behavior on a single example. Without reproducible interpretations, bias can only be measured---if at all---by studying the model's behavioral patterns across a whole evaluation set. 

In this situation, as opposed to Section \ref{sec:fixing-unwanted-behavior}, the model can produce plausible outputs using spurious correlations or shortcut learning. These heuristics are often undetectable in current evaluation practices. 
In \S\ref{sec:evals-ill-defined}, we will examine the case study of shortcut heuristics in Natural Language Inference (NLI). Then, in \S\ref{sec:counterfactual-interp}, we explore some counterfactual interpretability methods that can identify and validate underlying faulty reasoning patterns. 
Finally, in \S\ref{sec:robustness-yields-faithfulness}, we argue that reproducible and robust interpretability may ultimately yield faithfulness.

\subsection{Sometimes evaluative objectives are ill-defined}
\label{sec:evals-ill-defined}

In addition to invalid reasoning under specific regulatory requirements, some reasoning might be invalid because it fails to be reproducible under distribution shift. Previously, we discussed how to debug these errors after observation with falsifiable claims. However, these issues can be difficult to identify on in-distribution test sets; they are often discovered manually by chance or intuition. 
In  $\S\ref{sec:fixing-unwanted-behavior}$ we discussed how falsifiable interpretability may resolve errors discovered through standard evaluation practices. Here, we discuss how reproducible, \textit{counterfactual} interpretation may enable us to discover shortcuts and biases more elegantly, without blind experimentation.

One task in NLP where shifting evaluative goals and interpretability needs have intersected is natural language inference (NLI).
NLI is the task of predicting if a pair of statements logically entail or contradict each other. For a language model to handle this entailment task, it must learn natural language semantics \citep{merrill_can_2024}; in fact, the strength of next token predictors depends on their implicitly modeling entailment. Popular datasets for this task permitted models to rely on shortcut heuristics, allowing them to correctly predict the entailment of a sentence pair even when only one sentence was shown \citep{gururangan2018annotation,poliak2018hypothesis}. Because the task is defined in terms of both sentences, these heuristics would be brittle under some domain shifts. 

An array of interventions were proposed to remove these shortcuts in the models directly and in the datasets themselves. Model-in-the-loop dataset modification processes were used to either select new hard samples \citep{nie-etal-2020-adversarial} or filter shortcut-exhibiting samples \citep{saxon-etal-2023-peco}. At train time, residual learning \citep{he-etal-2019-unlearn}, adversarial training \citep{stacey-etal-2020-avoiding}, and automated counterfactual data augmentation \citep{wang2021counterfactual} all demonstrated that this ability can be unlearned or avoided during training altogether, even for models trained on biased datasets.

However, both this problem and its solutions were guided through experimentation and incremental changes to the evaluation objective by carefully selecting input examples. What if we could identify these reasoning flaws \textit{without manually creating counterfactual test sets}? 
If functional interpretability methods for debugging ($\S\ref{sec:fixing-unwanted-behavior}$) and detecting subtle failures ($\S\ref{sec:detecting-problems}$) can be applied to a trained NLI model, they could identify shortcut reasoning without human discovery, measure its use during inference, and efficiently patch the bias to convert a flawed model into a correct one.

The fundamental purpose of evaluation is to answer questions about what a model can do. Test sets are a necessary step along a claim-supporting chain of evidence for this purpose.
Poor within-task, cross-benchmark generalization is so well known that robustness to it is a means to benchmark base models \citep{yang-etal-2023-glue}. Models learning shortcuts and heuristics rather than generalizable mechanisms are an example of \textit{internal validity dangers}---benchmark-internal breaks in the chain of evidence \citep{liao2021we}.
These internal validity issues are often symptomatic of a problematic gap between the perception and truth of what an evaluation measures, or its \textit{construct validity} \citep{o1998empirical}. In the era of benchmarks attempting to measure abstract, generalized capabilities this problem has been exacerbated \citep{raji2021}.
A desire to measure generalized, abstract human-like cognition in models is natural in a field which aims to replicate human-like intelligence, although this goal is flawed in many ways \citep{Saxon2024BenchmarksAM}.
An evaluation's inability to map to a real-world task is a textbook \textit{external validity danger} \citep{liao2021we}.

\subsection{But counterfactual interpretation could help frame evaluative objectives}
\label{sec:counterfactual-interp}

Proactively identifying model internal vs. strictly externally diagnosable issues in evaluation would be a valuable goal for interpretability research. Instead of relying on manual efforts to mitigate problematic test set issues in \S\ref{sec:evals-ill-defined}, the most promising path forward lies in identifying shared mechanisms that, across tasks, degrade performance in complex ways that are understandable to humans. 
Interpretability work in the NLI bias explored which shortcut biases models learned during training and whether mitigation was necessary.
While NLI models are brittle under tests requiring non-heuristic syntactic \citep{mccoy-etal-2019-right} or lexical \citep{glockner-etal-2018-breaking} generalization, \citet{srikanth-rudinger-2022-partial} showed that models trained on the standard two-sentence task---even when exposed to shortcut features---do not necessarily rely on the single-sentence heuristics identified by \citet{poliak2018hypothesis} at test time.
This illustrates a broader point: evaluation alone often cannot settle debates about what models understand. Here, counterfactual interpretations that generalize across test distributions can step in where evaluations leave ambiguity. When evaluative objectives are underdefined, interpretability provides a way to probe whether a shared underlying mechanism governs performance across tasks.


When robust and trustworthy, counterfactual interpretability can help clarify whether performance across tasks reflects general capabilities or merely task-specific heuristics, especially when evaluation metrics alone are insufficiently precise. In this way, interpretability complements evaluation: evaluations elicit behaviors, and interpretability probes the causal mechanisms behind them. Thus, interpretability and evaluation are not separate tracks but complementary tools: evaluations act as diverse elicitation settings, while interpretability methods interrogate what is actually driving success. 

\subsection{Robust and reproducible explanations yield faithfulness in evaluation}
\label{sec:robustness-yields-faithfulness}

Without robustness, multiple such counterfactual interpretations can exist for a particular evaluation of behavior, each potentially revealing different results. Here, we argue an interpretation that satisfies robustness should help uncover causally faithful explanations--—interpretations that accurately reflect the model’s reasoning process \citep{wiegreffe-pinter-2019-attention, Jacovi2020TowardsFI, Lyu2022TowardsFM}—--even when the evaluative objective underlying observed behaviors is ill-defined. 
While the stability of interpretations could significantly vary depending on what criterion we use to measure faithfulness, robustness has shown to be quite effective \citep{Yin2021OnTF}. 
Although different interpretability methods serve distinct purposes and may highlight different aspects of the model, we should be able to trust their explanations if they are faithful under \textit{some} robustness guarantee.
From these faithful, robust explanations, we should then be able to establish insights that are commonly agreed upon--—some aspects of explanations should hold across different interpretations.
If robust interpretations are generally faithful, their insights should reveal common themes in the behavior being measured, even when evaluative objectives are ill-defined.
This, in turn, can help refine under-specified evaluative objective through robust explanations. \looseness-1

\subsection{From identification to intervention}
Falsifiability (\S\ref{sec:fixing-unwanted-behavior}) and reproducibility (\S\ref{sec:detecting-problems}) are minimal criteria for action, necessary but not sufficient. Our gender-MT example makes this concrete: both the profession-proxy and direct-gender readings satisfy reproducibility, yet only their differing OOD predictions can adjudicate which licenses an intervention. Predictivity (\S\ref{sec:predicting-problems}) is what closes this gap: a predictive interpretation specifies what should happen under counterfactual conditions, transforming a diagnosed mechanism into one we can act on.

\section{Predicting future failures}
\label{sec:predicting-problems}

Beyond the two previously discussed scenarios---debugging failures on known evaluative objectives (\S\ref{sec:fixing-unwanted-behavior}) and detecting subtle failures under ill-defined evaluative objectives (\S\ref{sec:detecting-problems})---there are cases in which the evaluation objectives themselves are unknown. In \S\ref{sec:predicting-problems}, we explore a third scenario: potential model errors that go unnoticed because existing evaluations fail to capture the full range of data conditions and failure modes.
We argue that if interpretability can produce predictive claims, we may use them to anticipate model failures from internal signals. Specifically, we may generate evaluations that are sufficiently challenging, tailored to the model.
By systematically exploring and formalizing criteria for problematic behaviors or challenging inputs, interpretability facilitates a proactive strategy for uncovering model limitations. 

A clarification is necessary about what we mean by ``prediction." In machine learning, ``prediction" refers to a model's output: given an input, the system produces a label, token, or score. This notion of prediction is in some sense purely behavioral and operational. By contrast, in the scientific sense, prediction plays a fundamentally different role. A scientific theory and hypothesis demonstrates its depth by its testable expectations about what should occur under sufficiently specified assumptions and conditions. The success or failure of these predictions becomes evidence for---or against---the adequacy of our understanding. Under this framing, prediction becomes the strongest form of evaluation: rather than retrospectively summarizing performance, we generate mechanism-grounded hypotheses and design evaluations to validate our understanding.\looseness-1

\paragraph{Extending the gender-translation example} Consider a hypothetical scenario in which no researcher has yet discovered gender bias in machine translation; for example, the possibility of conflating profession and gender has never been considered.
Without exhaustively enumerating evaluation scenarios, we could instead examine model internal responses to the training data to anticipate likely failure cases.
We may leverage interpretability to discover geometric entanglements between profession- and gender-related features, with which we generate evaluation sets that the models may struggle to generalize when associations from the training data no longer hold. 
Under sufficient robustness assumptions and conditions, we could leverage that very geometric properties of these relations to propose this form of bias as an evaluative objective even if we never encounter a female doctor in the existing data.

We propose that, if we framed interpretability claims in terms of scientific hypotheses to generate predictions about model behaviors, predictive interpretability can be leveraged to design evaluative objectives that specifically target areas where the model is likely to underperform. We begin by examining how geometric properties identifiable by interpretability can support predictive evaluation (\S\ref{sec:rep-and-geo-properties}). This motivates the need for a precise understanding of spurious correlations and their relationship to model internals (\S\ref{sec:spurious-correlation}). Finally, we explore how to construct such evaluation sets by employing mechanisms for predicting out-of-distribution (OOD) behavior (\S\ref{sec:ood}).

\subsection{We can use geometric properties of representations for \textit{prediction}}
\label{sec:rep-and-geo-properties}

Models learn geometric representations that reflect patterns in the training data, often aligning with how humans intuitively relate certain concepts. 
The classic example is the embedding vector relation ``man + woman = king + queen,'' 
which captures gendered relationships in a way that mirrors human intuition \citep{Vylomova2015TakeAT, Liu2022AreRB}.
Earlier work demonstrated that word embeddings can encode female and male gender stereotypes along a specific direction in the embedding space \citep{Bolukbasi2016ManIT}. This insight has already been used to address issues in model behavior: understanding the geometry of bias enables targeted interventions in embedding spaces for debiasing \citep{Kaneko2021DebiasingPC, Gonen2019LipstickOA}. \looseness-1

Recently, the geometry of learned representations has been further elucidated, characterized in terms of the intrinsic dimensionality of multi-head attention embeddings and identified per-layer affine mappings of feedforward networks, facilitating a deeper understanding of model behavior in tasks like toxicity detection \citep{Balestriero2023CharacterizingLL}. These examples illustrate how uncovering the geometric structure of representations can not only reveal problematic model behaviors but also suggest it provides avenues for proactively defining and predicting them.

These geometric properties are not only observable but also mathematically definable and structurally persistent \citep{Gardinazzi2024PersistentTF, Park2023TheLR}. Another significant recent development is the linear representation hypothesis \citep{Elhage2022ToyMO, Park2023TheLR, Li2022EmergentWR}, which posits that high-level concepts are encoded as linear directions in representation space. 
However, this hypothesis may break down for significantly out-of-domain data \citep{strongfeature2024}, where robustness failures undermine linear representation. Consider the most extreme scenario, in which we identify a feature that appears to represent gender, but in fact no longer corresponds to gender at all when the context relates to individual profession. In such a situation, even if we have interpreted features representing gender and profession, we cannot predict model behavior from their interaction because their interpretations do not hold in combination.
If we aim to use these mathematical definitions of representations to define evaluative objectives and guide generations of challenge sets, we therefore want robustness guarantees.

Operationalizing this claim requires committing to a specific geometric quantity. Candidates include subspace alignment angles, intrinsic dimensionality, or distributional distances between concept subspaces; each making a different falsifiable prediction about OOD failure. We do not advocate a single choice, but observe that the geometric argument is only as predictive as the quantity it commits to. 
Geometric analysis also inherits the underdetermination we diagnosed for SAEs in \S\ref{sec:disconnect_interp_eval}: rank collapse and manifold shift are consistent with the same failure pattern, so predictivity requires not just a geometric quantity but a discriminating one.

\subsection{To use interpretation for prediction, we should understand spurious correlations}
\label{sec:spurious-correlation}

In embedding space, concepts exhibit geometric properties that allow certain representations to be combined or separated to form other intuitive concepts \citep{Lappin2022ANM, Wattenberg2024RelationalCI, Lepori2023BreakID}. 
This structure not only enables interpretability but also helps predict and mitigate biases or other problematic behaviors. However, even if geometric interpretations are stable, a key challenge remains: models often learn spurious correlations that may not be evident through geometric analysis alone.

Detecting these spurious correlations 
remains a significant challenge for current interpretability methods. 
SAEs have shown promise in low-data or corrupted-data settings—particularly when spurious features are simple—but often fail on OOD data, frequently returning null results \citep{kantamneni2024sae, Karvonen2024EvaluatingSA}. 
When effective, SAEs may still help uncover dataset defects or mislabeled examples \citep{kantamneni2024sae}. 
Moreover, spurious correlations might pose a more fundamental challenge.
Prior work suggests that spurious correlations may reflect deeper limitations of Empirical Risk Minimization (ERM), which can lead to ``causally confused'' models that overfit to patterns that generalize poorly—particularly on OOD inputs \citep{causal_confusion_scaling, Krueger2020OutofDistributionGV}. 
Since ERM minimizes average risk, models may exploit spurious features to reduce training loss---an issue not easily resolved through scaling or naive fine-tuning. 
Even increasing data diversity may be insufficient, as concept shifts driven by unobserved causal factors can persist in large datasets \citep{Krueger2020OutofDistributionGV}.

\subsection{Predicting out-of-domain behavior with in-domain interpretability}
\label{sec:ood}

Full understanding of a model through interpretability may enable us predict how it behaves with OOD data \citep{Juneja2022LinearCR}. 
Estimating performance in OOD scenarios, especially where labeled data is scarce, is critical for safe deployment. A deeper understanding of how models generalize across distributions allows us to anticipate their behavior on unfamiliar inputs. 
In this section, we explore analyses that can be extended to predict evaluative criteria and generate targeted OOD test cases.

Detecting generalization failures remains a significant challenge for current interpretability methods. 
At model internal level, predicting OOD behavior may be possible by leveraging various forms of model invariance. Prior work has shown that such invariances can improve generalization under distribution shifts, including domain changes \citep{Gulrajani2020InSO}, causal interventions \citep{Arjovsky2019InvariantRM}, data augmentation \citep{Cubuk2019RandaugmentPA}, and local interpolations \citep{Luo2017SmoothNO}. Models invariant to local transformations tend to factorize input space into a base space and a transformation set, effectively reducing input dimensionality and model complexity—thereby enhancing generalization. One example is neighborhood invariance, a complexity measure for learned representations. \citeauthor{Ng2022PredictingOG} showed that neighborhood invariance remains robust even in OOD settings where other methods fail. Because it only requires selecting appropriate data transformations, this approach presents a promising direction for extending model performance beyond the training distribution.\looseness=-1

\section{Conclusion}

\begin{table}[h]
\centering
\small
\renewcommand{\arraystretch}{1.2}
\setlength{\tabcolsep}{4pt}
\begin{tabular}{@{}lccc@{}}
\toprule
\textbf{} & \textbf{Falsifiability} & \textbf{Reproducibility} & \textbf{Predictability} \\
\midrule
SAEs      & \prt{} & \no{}  & \prt{} \\
CBMs      & \yes{} & \prt{} & \no{}  \\
Attention & \no{}  & \no{}  & \no{}  \\
Probing   & \prt{} & \yes{} & \prt{} \\
\bottomrule
\end{tabular}
\caption{Interpretability against scientific standards. }
\label{tab:methods}
\end{table}

In sum, behavioral benchmarks tell us what models do, but not how or why. Interpretability, when held to scientific standards of falsifiability, reproducibility, and predictability, offers a path toward mechanism-level evaluation that extends beyond traditional metrics. Table~\ref{tab:methods} makes the gap concrete: across four common method families, none currently meets all three criteria. By enabling causal debugging, reliable detection of faulty reasoning, and anticipation of future failures, interpretability can transform evaluation from a surface-level score-keeping into a scientifically grounded explanation of internal process. If developed rigorously, interpretability can become evaluation in its own right. \looseness-1

\section*{Acknowledgements}
IL is supported by Coefficient Giving's Technical AI Safety Research Grant. 
We are grateful to Naomi Saphra for insightful discussions and detailed comments.

\bibliography{references}
\end{document}